\documentclass{jetpl}

\usepackage{graphicx}
\usepackage{cite}

\twocolumn
\lat

\renewcommand{\Re}{\mathop{\rm Re}}
\renewcommand{\Im}{\mathop{\rm Im}}

\newcommand{\eps}{\varepsilon}

\newcommand{\vvr}{{\bf r}}
\newcommand{\vH}{{\bf H}}
\newcommand{\vA}{{\bf A}}

\newcommand{\dd}{{\rm d}}
\def\be{\begin{equation}}
\def\ee{\end{equation}}

\begin{document}

\title{On the Effect of Far Impurities on the Density of States
of Two-Dimensional Electron Gas in a Strong Magnetic Field}
\rtitle{On the effect of far impurities on the density of states
of 2DEG in a strong magnetic field}
\sodtitle{On the Effect of Far Impurities on the Density of States
of Two-Dimensional Electron Gas in a Strong Magnetic Field}

\author{I.\,S.\,Burmistrov$^{*+}$\/\thanks{e-mail: burmi@itp.ac.ru, skvor@itp.ac.ru}
                                                      and M.\,A.\,Skvortsov$^{*}$}
\rauthor{I.\,S.\,Burmistrov and M.\,A.\,Skvortsov}
\sodauthor{Burmistrov, Skvortsov}


\address{$^{*}$L.D. Landau Institute for Theoretical Physics,
Kosygina str. 2, 117940 Moscow, Russia\\
$^{+}$Institute for Theoretical Physics, University of Amsterdam,
Valckenierstraat 65, 1018 XE Amsterdam, The Netherlands}

\abstract{The effect of impurities situated at different
distances from a two-dimensional electron gas on the density of
states in a strong magnetic field is analyzed.
Based on the exact result of Brezin, Gross, and Itzykson,
we calculate the density of states in the whole energy range,
assuming the Poisson distribution of impurities in the bulk.
It is shown that in the case of small impurity concentration the
density of states is qualitatively different from the model case
when all impurities are located in the plane of the two-dimensional
electron gas.}


\maketitle




{\bf 1. Introduction.}
Two-dimensional electrons in a quantizing magnetic field $H$
has been attracting much attention~\cite{AFS} especially since the
discovery of the quantum Hall effect~\cite{QHE}. The properties of
two-dimensional electrons in the magnetic field are affected by
the presence of electron-electron interaction as well as by
impurities. Investigation of the density of states as a function
of the magnetic field and filling fraction allows us to estimate
the inhomogeneities caused by impurities in experimental
samples~\cite{KMT}. Although the electron-electron interaction
should usually be taken into account, the question about the
density of states in the simplest model of noninteracting
electrons is also rather interesting.

In the absence of interaction, impurities near a two-dimensional
electron gas (2DEG) provide the only mechanism for broadening of
Landau levels.
In a weak magnetic field the large number of Landau levels, $N\gg
1$, are filled. One can therefore use the self-consistent Born
approximation that is justified by the small parameter $\ln N/N\ll
1$. It results in the well-known semicircle shape for the density
of states~\cite{Ando}. Beyond the self-consistent Born
approximation one can find the exponentially small tails in the
density of states~\cite{EM}.

In the opposite limit of a strong magnetic field only the lowest
Landau level is partially occupied. In this case one can neglect
the influence of the other empty Landau levels assuming $\omega_H
\gg T, \tau^{-1}$. Here $\omega_H=e H/m$ denotes the cyclotron
frequency with $e$ and $m$ being the electron charge and mass
respectively, $T$ stands for the temperature, and $\tau$ is the
elastic collision time. The density of states on the lowest Landau
level strongly depends on the statistic properties of the random
potential created by impurities and on the value of the
dimensionless parameter $n_S/n_L$, where $n_L=1/(2\pi l_H^2)$ with
the magnetic field length $l_H=1/\sqrt{m \omega_H}$ and $n_S$
stands for the two-dimensional impurity density. For the
white-noise distribution of the random potential the density of
states was found exactly by Wegner~\cite{W}. For arbitrary
statistics of the random potential the density of states was
obtained exactly in a beautiful paper by Brezin, Gross, and
Itzykson~\cite{BGI}. If the number of impurities is less than the
number of states on the Landau level, $n_S < n_L$, the Landau
level remains partially degenerate. In the opposite case $n_S \geq
n_L$ the presence of impurities leads to complete lifting the
degeneracy of the Landau level~\cite{BME, BGI}.

In experimental samples the impurities can be found rather far
from the 2DEG~\cite{AFS,QHE}. In such a situation the two-dimensional
electron system is subject to the three-dimensional random potential.
It means that an electron localized at the heterojunction feels
impurities situated at distances much larger than the width $z_0$
of the 2DEG.
This situation was considered recently by Dyugaev, Grigor'ev,
and Ovchinnikov~\cite{DGO}. Within the lowest order of the
perturbation theory in the concentration $n_{\rm imp}$ of
three-dimensional scatterers,
they have calculated the density of states $D(E)$ in the limit
when the multiple scattering on the same impurity provides
the main contribution. Assuming exponential decay of the wave function
in the transverse direction, $\varphi^2(z) \propto \exp(-z/z_0)$,
they obtained a universal regime where $D(E)=n_{\rm imp}z_0/E$,
and energy $E$ is measured from the unperturbed Landau level.
Being bounded both from the sides of small and large energies
by many-impurity effects, this interval contains most of the states
of the unperturbed Landau level.
Though the analysis of Ref.~\cite{DGO} holds for an arbitrary Landau
level, it cannot be generalized to the limits of small and large energies
where a nonperturbative treatment of impurity scattering is required.

The main objective of the present letter is to present the full
analysis of the effect of far impurities on the density of states
of a two-dimensional electron gas in a strong magnetic field.
Employing the remarkable result of Brezin, Gross, and
Itzykson~\cite{BGI} we calculate the broadening of the lowest Landau
level by the three-dimensional short-range impurities with the Poisson
distribution in the bulk.

{\bf 2. Results.}
Usually impurities occupy rather large volume near a two-dimensional
electron gas and, consequently, their number exceeds the number of
states on the Landau level, $N_{\rm imp} \gg n_L S$,
with $S$ being the area of two-dimensional electron system.
Therefore,
the degeneracy of the Landau level is removed completely by
impurities~\cite{BGI, DGO}. The behavior of the density of states
is determined by the new dimensionless parameter
\be
\label{f}
  f = \frac{n_{\rm imp} z_0}{n_L} ,
\ee
that will be referred to as impurity concentration.
Here $n_{\rm imp}$ is the three-dimensional impurity density and $z_0$
stands for the spatial extent of the electron wave function in the
direction perpendicular to the 2DEG,
explicitly defined in Eq.~(\ref{varphi}).

In experiments there is usually a small amount of impurities in a
layer of width $z_0$ near the two-dimensional electron gas~\cite{AFS,QHE},
i.e. impurity concentration is small, $f\ll 1$.
In this case we obtain the following density of states
at the lowest Landau level as a function of the deviation $E$
from the unperturbed level $\omega_H/2$:
\begin{equation}
\label{DOSRES}
\frac{D(E)}{n_L} =
  \left \{ \!
  \begin{array}{lr}
    0 , & E < 0,\\
    \displaystyle \frac{1}{E_0} S_f \Bigl( f\ln\frac{E_0}{E} \Bigr) , &
        \displaystyle 0\leq \frac{E}{E_0} \ll e^{-1/f},\\
    \displaystyle \frac{f}{E} , &
      \displaystyle e^{-1/f}\ll \frac{E}{E_0} \ll 1, \\
    \displaystyle \frac{2\pi^{-1/2} E^2}{(f_1 E_1^2)^{3/2}}
      \exp \Bigl ( \lefteqn{ -\frac{E^2}{f_1 E_1^2}\Bigr ) ,} &
        E_0 \ll E,
  \end{array}\right .
\end{equation}
where
\be
\label{Sf}
  S_f(\xi) =
  \sqrt\frac{f}{2\pi} \frac{\xi-1}{\Gamma(\xi)}
  \exp\Bigl ( \frac\xi f - \frac{(\xi-1)^2}{2f}- \frac{f}{2}\ln^2 \xi\Bigr) .
\ee
On deriving Eq.~(\ref{DOSRES}) we have assumed that the wave function
$\varphi(z)$ decays in the transverse direction as
\begin{equation}
\label{varphi}
  \varphi^2(z) \sim \frac{A}{z_0}\exp \left (-\frac{z}{z_0}\right),
    \qquad
  z\gg z_0,
\end{equation}
with $z_0$ defining the width of the 2DEG,
and $A$ being a constant of order 1.
The form (\ref{varphi}) corresponds to a rectangular well confining
potential~\cite{AFS}.
The energy scale
\be
\label{E0-def}
  E_0 = e^\gamma A \, \frac{n_L u_0}{z_0}
\ee
is introduced by impurities, with $u_0>0$
being the strength of the repulsive disorder potential
[cf.\ Eq.~(\ref{V1}) below],
and $\gamma\approx 0.577$ denoting the Euler constant.
The result for $E\gg E_0$ is governed by the parameters
\be
\label{f1}
  f_1 = \frac{n_{\rm imp} z_1}{n_L} \sim f ,
\qquad
  E_1 = \frac{n_L u_0}{z_1} \sim E_0 ,
\ee
where $z_1 \sim z_0$ is the width of the wave function as
determined via its fourth moment:
\be
\label{z1}
  \frac{1}{z_1} = \int_0^\infty \varphi^4(z) dz .
\ee

The fact that the density of states vanishes for $E<0$ is expected
since the random potential is purely repulsive.
Since $\lim_{\xi\to\infty}S_f(\xi)=0$, the density of states also
vanishes at the position of the unperturbed Landau level, $D(0)=0$.

In the interval $0\leq E\ll E_0 e^{-1/f}$ the density of states
exhibits the maximum
\be
\label{rhomax}
  D_{\rm max}(E_*) = \sqrt{\frac{f}{2\pi}}\frac{n_L}{E_0}
  \exp\Bigl [ \frac{3}{2f} \Bigr ]
\ee
at the exponentially small energy
\be
\label{Emax}
  E_{*} \sim E_0 \exp\Bigl(-\frac 2 f \Bigr ).
\ee

In the region $E_0 e^{-1/f}\ll E \ll E_0$ the density of states
$D(E)=fn_L/E$ is linear in impurity concentration coinciding
with the perturbative result obtained by
Dyugaev, Grigor'ev and Ovchinnikov~\cite{DGO}. It indicates that
the multiple scattering on the same impurity provides the main
contribution to the density of states for energies $E_0
e^{-1/f}\ll E \ll E_0$. This energy interval contains the major
part of the states formed from the lowest Landau level.

The result announced in Eq.~(\ref{DOSRES}) in the limit
$0\leq E\ll E_0 e^{-1/f}$ is applicable for
$f\ln(E_0/E)-1\gg\sqrt{f}$, cf.\ Eq.~(\ref{ReF}).
At the border of applicability Eq.~(\ref{Sf}) gives
$S_f[f\ln(E_0/E)] \sim fE_0/E$, and thus $D(E)$ merges
with the universal result at $E\gg E_0 e^{-1/f}$.

In the region of rather large energies $E \gg E_0$, the tail of
the density of states is described by the same expression as if all
impurities were situated in the plane of the 2DEG,
with the effective two-dimensional parameters
\begin{equation}
\label{eff}
  u_0^{(2D)} = \frac{u_0}{z_1},
\qquad
  n_{\rm imp}^{(2D)} = n_{\rm imp} z_1.
\end{equation}
We mention that the tail of the density of states corresponds to
some optimal fluctuation of the random potential as it happens for
the pure two-dimensional problem~\cite{IL, A}.

For large impurity concentration, $f \gg 1$, the Poisson
distribution can be replaced by the white-noise distribution of
impurities on the plane with the effective parameters \eqref{eff}.
The density of states is given therefore by the well-known
formula~\cite{W,BGI}
\begin{equation}
\label{DOSRES2}
  D(E) = \frac{n_L}{\pi E_1\sqrt{f_1} } W\left ( \frac{E-f_1
E_1}{E_1\sqrt{f_1} }\right ),
\end{equation}
where  we introduce the function
\begin{equation}\label{W}
W(z) =  \frac{2}{\sqrt\pi} e^{z^2}\left [ 1+ \frac{4}{\pi}\Bigl
[\int \limits_{0}^{z}e^{x^2}\dd x\Bigr ]^2\right ]^{-1}.
\end{equation}
The shift of the maximum of $D(E)$ to positive energies
is related to repulsive character of impurities' potential.
Eq.~\eqref{DOSRES2} describes the density of states for the Poisson
distribution only approximately, since the exact density of states
should vanish $E\leq 0$.
However, deviation of Eq.~\eqref{DOSRES2} from the exact
answer is exponentially small ($e^{-f}\ll 1$) for positive $E$.


{\bf 3. The model.}
The spin-polarized two-dimensional electron gas in the presence of
the random potential $V(\vvr,z)$ and the strong perpendicular
magnetic field $\vH$ is described by the following one-particle
Hamiltonian
\begin{equation}\label{H1}
{\cal H} = -\frac{1}{2 m}\left (\nabla - i e\vA \right )^2 +
V(\vvr,z)+ U_{\rm conf}(\vvr,z) .
\end{equation}
Here $\vA$ stands for the vector potential, $\vH = {\rm rot} \, \vA$
and $U_{\rm conf}(\vvr,z)$ denotes the confining potential that
creates the two-dimensional electron gas. We use the units such
that $\hbar=1$ and $c=1$.

We assume that impurities situated near the two-dimensional
electron gas are zero-range repulsive ($u_0>0$) scatterers
producing the random potential
\begin{equation}\label{V1}
V(\vvr) = u_{0} \sum \limits_{j=1}^{N_{\rm
imp}}\delta^{(2)}(\vvr-\vvr_{j})\delta(z-z_{j}).
\end{equation}

Assuming that the confining potential $U_{\rm conf}$ depends only
on the $z$ coordinate,  we can represent the electron wave
function as follows
\begin{equation}\label{Psi}
\Psi(\vvr,z) = \psi(\vvr) \varphi(z),
\end{equation}
where $\varphi(z)$ is the ground-state wave function for the
electron motion in the direction perpendicular to the 2DEG in the
absence of disorder, and $\psi(\vvr)$ describes the electron
motion in the plane of 2DEG. The decomposition (\ref{Psi}) is
equivalent to the projection onto the lowest level of dimensional
quantization and is analogous to the projection onto the lowest
Landau level states $\psi(\vvr)$. Since in experiment the energy
separation between the lowest and the first excited level of
dimensional quantization is usually larger than the cyclotron gap,
the accuracy of projection onto $\varphi(z)$ is higher than the
accuracy of projection onto the lowest Landau level. With the help
of the ansatz (\ref{Psi}) the original three-dimensional problem
\eqref{H1} reduces to the two-dimensional one with the effective
two-dimensional random potential
\begin{equation}\label{V2}
V_{\rm eff}(\vvr) = u_{0} \sum \limits_{j=1}^{N_{\rm imp}}
\varphi^2(z_j) \delta^{(2)}(\vvr-\vvr_{j}).
\end{equation}
Thus, the distribution of impurities along the $z$ direction
leads to an additional random distribution of the potential strengths
$u_{0} \varphi^2(z_j)$ effectively felt by two-dimensional electrons.

By using the general result of Brezin, Gross, and Itzykson~\cite{BGI}
for the random potential \eqref{V2}, we obtain for the density of states
at the lowest Landau level:
\begin{equation}
\label{DOS1}
  D(E) = \frac{n_L}{\pi} \Im \frac{\partial}{\partial E} \ln F(E),
\end{equation}
where
\begin{equation}
\label{F1}
F(E) = \int \limits_{0}^{\infty} \dd t \exp  \left (\frac{i E
t}{n_L} + \int \limits_{0}^{t} \frac{\dd \beta}{\beta}
g(\beta)\right).
\end{equation}
The properties of the random potential are encoded
in the function $g(\beta)$ which is defined as
\begin{equation}\label{g0}
  \exp \Bigl\{ n_L \! \int \dd^{2} \vvr g[\beta(\vvr)] \Bigr\}
  =
  \left \langle \exp
    \Bigl [- i \int \dd^{2}\vvr \beta(\vvr) V_{\rm eff}(\vvr) \Bigr]
  \right \rangle,
\end{equation}
where the average $\langle \cdots \rangle$ is with respect to the
distribution of the random potential $V_{\rm eff}(\vvr)$.

We assume that the three-dimensional scatterers (\ref{V1}) with equal
strengths $u_0$ obey the Poisson statistics, being uniformly distributed
along the $z$ direction. Then averaging over $V_{\rm eff}(\vvr)$
in Eq.~(\ref{V2}) reduces to integration over $z$ coordinate:
\be
\label{g1}
g(\beta) = \frac{n_{\rm imp}}{n_L} \int \limits_{0}^{\infty}\dd z
\left (e^{ -i \beta u_0 \varphi^2(z)}
 -1 \right ).
\ee
On writing Eq.~(\ref{g1}) we employed the fact that the wave function
$\varphi(z)$ vanishes for $z<0$.

{\bf 4. Evaluation of the density of states.}
The density of states is generally given by the integral representation
(\ref{DOS1}), (\ref{F1}) and (\ref{g1}). However, Eq.~(\ref{F1}) cannot
be calculated in a closed form valid for arbitrary values of impurity
concentration and energies. Below we analyze the most interesting
asymptotic cases.

First of all, we note that $D(E)$
vanishes for energies $E<0$ regardless of the form of $\varphi(z)$.
This follows from the fact that for $E<0$ the function $F(E)$
is purely imaginary that can be obtained by performing the Wick
rotation $t\to -i\tau$ of the integration contour in Eq.~(\ref{F1}).

The density of states can also be easily calculated in the limit
of either large impurity concentration ($f\gg1$) and arbitrary energies,
or small impurity concentration ($f\ll1$) but large energies $E\gg E_0$.
In both cases the integral (\ref{F1}) is determined by small values of $t$
that allows to expand the function $g(\beta)$ given by Eq.~(\ref{g1}):
\be
\label{htSer}
  \int \limits_{0}^{t} \frac{\dd \beta}{\beta} g(\beta)
  \simeq
  - i \frac{n_{\rm imp} u_0 t}{n_L}
  - \frac{n_{\rm imp} u_0^2 t^2}{4n_Lz_1} ,
\ee
where $z_1$ is defined in Eq.~(\ref{z1}).
The quadratic term in Eq.~(\ref{htSer}) describes Gaussian (white-noise)
distribution of impurities~\cite{BGI}, whereas the linear term
accounts for the energy shift due to the nonzero average potential
of impurities. Employing the result of Ref.~\cite{BGI}, we arrive at
Eq.~(\ref{DOSRES2}).
Using the asymptotic expression
$W(x)\simeq 2 \sqrt\pi x^2 e^{-x^2}$ valid at $x\gg 1$,
we obtain the result \eqref{DOSRES} for $f\ll1$ in the regime $E \gg E_0$.

The most interesting is the behavior of $D(E)$ in the limit of
{\em small impurity concentrations}, $f\ll1$, and {\em
sufficiently small energies}, $E\ll E_0$. In this limit assumed
hereafter the function $F(E)$ given by Eq.~(\ref{F1}) is
determined by large values of $t$ that allows to use the
asymptotic formula (\ref{varphi}) for calculation of $g(\beta)$ in
Eq.~(\ref{g1}). Introducing the dimensionless energy $\eps =
E/E_0$ where the energy scale $E_0$ is defined in
Eq.~(\ref{E0-def}) and rescaling $t$ accordingly we rewrite the
expression for the density of states as
\begin{equation}
\label{DOS2}
  D(\eps) = \frac{n_L}{\pi E_0} \Im \frac{\partial}{\partial \eps}
\ln F(\eps),
\end{equation}
where
\begin{gather}
\label{F2}
F(\eps) = \int \limits_{0}^{\infty} \dd t e^{ie^{\gamma}\eps t} e^{ - f h(t)},
\\
\label{h1}
h(t) = \int \limits_{0}^{t} \frac{\dd \beta}{\beta} \int
\limits_{0}^{\beta} \frac{\dd s}{s} \left (1 - e^{-is}\right).
\end{gather}

The function $h(t)$ is positive at the negative part of the
imaginary axis, $t=-i\tau$, having the following asymptotic behavior
at $\tau\gg1$:
\be
  h(-i\tau) = \frac12 \ln^2 (e^\gamma\tau) + c_0 + \hat h(\tau) ,
\ee
where $c_0$ is a constant of the order 1,
and $\hat h(\tau)$ decays exponentially at large $\tau$:
\be
  \hat h(\tau) = - \int_1^\infty \frac{dx}{x} e^{-\tau x} \ln x
  \approx -\frac{1}{\tau^2} e^{-\tau} .
\ee
The $\ln^2 t$ asymptotics of $h(t)$ is specific to the problem
with distributed strengths $u_0\varphi^2(z_j)$ of impurities
and asymptotic behavior (\ref{varphi}) of the wave function
$\varphi(z)$ far from the 2DEG,
and should be contrasted with the $\ln t$ dependence for the case
of the Poisson distribution with constant impurities' strengths.
For another decay law of the wave function,
$\varphi^2(z) \sim \exp[-(z/z_0)^\alpha]$,
the leading asymptotics would be $h(t) \sim \ln^{1+1/\alpha}t$.

The function $F(\eps)$ in Eq.~(\ref{F2}) is given by an oscillating
integral. Therefore it is desirable to deform the integration contour
to get rid of oscillations. However, for $\eps>0$ such a deformation
in Eq.~(\ref{F2}) is impossible: the first factor prohibits deformation
into the lower half-plane, whereas the second factor leads to a divergent
integral if deformed  into the upper half-plane.
This complication can be overcome by splitting the integrand into two
parts singling out the leading log-square asymptotics:
\be
\label{F3}
  F(\eps) = \int \limits_{0}^{\infty} \dd t
  e^{ie^{\gamma}\eps t - (f/2) \ln^2 (e^\gamma it)}
  \bigl[ 1 + (e^{-f\hat h(it)}-1) \bigr],
\ee
where we have omitted an irrelevant factor $e^{-fc_0}$.
In the limit $\eps<1$, the integral with the second term
($e^{-f\hat h(it)}-1$) in the square brackets allows deformation
of the contour to the negative part of the imaginary axis,
where the integrand is purely real. It can be shown that
the resulting contribution can be neglected compared to the
integral with the first term in the square brackets.
The latter can be calculated by deforming the integration contour
to the upper part of the imaginary axis.
After a proper rescaling of variables one finds:
\be
\label{F4}
  F(\eps) =
  \int \limits_{0}^{\infty} \dd \tau
  \exp\left(
     -\tau - \frac{f}{2}\bigl[\ln \frac{\tau}{\eps}+i\pi\bigr]^2
  \right),
\ee
where we have again omitted an irrelevant factor $ie^{-\gamma}/\eps$.

Equations (\ref{DOS2}) and (\ref{F4}) give the integral
representation for the density of states at $E\ll E_0$.
Its behavior depends on the value of the parameter
\be
\label{xi}
  \xi=f\ln \frac{1}{\eps} .
\ee

{\em For small $\xi<1$}, i.e., not too close to the unperturbed Landau level
($e^{-1/f}\ll\eps\ll1$), one can calculate $F(\eps)$ perturbatively.
Expanding Eq.~(\ref{F4}) in $f$ one can easily recover the perturbative
result of Ref.~\cite{DGO} as well as the leading correction to it:
\begin{equation}
\label{DOS5}
  D(\eps) = \frac{n_Lf}{E_0\eps}
    \left [
      1
      - 2\zeta(3) f^2 \ln\frac1\eps + \dots
    \right ] .
\end{equation}
Retaining only the leading term we obtain the result~\eqref{DOSRES}
in the regime $E_0 e^{-1/f}\ll E \ll E_0$.

{\em For large $\xi>1$}\/ corresponding to  energies close
to the unperturbed Landau level, evaluation
of Eq.~(\ref{F4}) is subtler. In this case the ratio
$\Im F(\eps)/\Re F(\eps)$ is exponentially small and special care
must be taken in order to extract $\Im F(\eps)$.
On the other hand, $\Re F(\eps)$ can easily be calculated
for $\xi-1\gg\sqrt{f}$. Making the substitution $\tau=\eps e^p$
and calculating the resulting Gaussian integral over $p$
one finds
\be
\label{ReF}
  \Re F(\eps) =
  - \eps \sqrt{\frac{2\pi}{f}}
    \exp \left[ \frac{1}{2f} \right] ,
\qquad
  \xi-1\gg\sqrt{f}.
\ee
To extract $\Im F(\eps)$, we find it convenient to pass to another
representation for the function $F(\eps)$. To this end we
decouple the square term in the exponential of Eq.~\eqref{F4}
by the Hubbard-Stratonovich transformation, and integrating over $\tau$
we obtain
\be
\label{F12}
  F(\eps) = \int \limits_{-\infty}^{\infty} \frac{dz}{\sqrt{2 \pi f}}
  \: \Gamma(1+i z)
  \exp\Bigl [ -\frac{z^2}{2f} -\pi z + i z \ln \frac{1}{\eps}\Bigr ].
\ee
This representation in terms of the $\Gamma$-function is suitable
for numerical simulation due to rather fast convergence of the integral,
contrary to the initial representation ({\ref{F2}}).

To proceed we shift the integration contour to the upper part of the complex
plane: $z=i\xi+x$, with $x$ being the new real integration variable.
As soon as $\xi \geq 1$, in doing the contour transformation,
we have to cross the poles of the $\Gamma$-function at $z=ik$ with
integer $k>0$. As a result we obtain
\begin{gather}
F(\eps) = \frac{1}{\sqrt{2\pi f}}
    \Biggl\{
      2\pi \sum_{k=1}^{[\xi]} \frac{(-1)^{k}}{(k-1)!}
        \exp\Bigl [ \frac{k^2-2 \xi k}{2f}\Bigr ]
\nonumber \\
{}      + \exp \Bigl (-\frac{\xi^2}{2f} \Bigr ) \Phi(\xi,f)
    \Biggr\},
\label{Ph1}
\end{gather}
where $[\xi]$ is an integer part of $\xi$, and the function
$\Phi(\xi,f)$ is defined as
\begin{equation}\label{DefPh}
\Phi(\xi,f)=e^{-i\pi\xi}\int \limits_{-\infty}^{\infty} \dd x \,
\Gamma(1-\xi+ix)e^{-\pi x}
        \exp\Bigl [ -\frac{x^2}{2f} \Bigr ].
\end{equation}

An advantage of this representation is that the pole contribution in
Eq.~(\ref{Ph1}) is purely real and hence $\Im F(\eps)$ is determined
solely by $\Im\Phi(\xi,f)$. Employing the identity
$\Gamma(1-\eta)\Gamma(\eta)=\pi/\sin(\pi\eta)$ with $\eta=\xi-ix$
we obtain for the imaginary part of $\Phi(\xi, f)$:
\begin{equation}
\label{Ph4}
  \Im \Phi(\xi,f)=-2 \pi \Re
    \int\limits_{0}^{\infty} \dd x \,
    \frac{\displaystyle\exp\Bigl [ -\frac{x^2}{2 f}\Bigr ]}{\Gamma(\xi-i x)}.
\end{equation}
In the limit $\xi \gg 1$, the term $ix$ in the argument of the
$\Gamma$-function can be taken into account as
$\Gamma(\xi-ix) \simeq \Gamma(\xi) e^{-i x \ln \xi}$.
Thereby we find the following estimate:
\begin{equation}
\label{ImPhi}
  \Im \Phi(\xi,f) = -\frac{\pi \sqrt{2\pi f}}{\Gamma(\xi)}
  \exp\Bigl [ -\frac{f}{2}\ln^2\xi\Bigr ].
\end{equation}
Though Eq.~(\ref{ImPhi}) is formally derived for $\xi\gg1$,
it can also be applied at $\xi\geq1$ as well, with the error being
small in virtue of the inequality $f\ll1$.

Now with the help of Eqs.~\eqref{ReF}, \eqref{Ph1} and \eqref{ImPhi}
we obtain for $\xi-1 \gg \sqrt{f}$:
\begin{gather}
\label{F21}
  \frac{F(\eps)}{\Re F(\eps)} =
  1 + i\sqrt{\frac{\pi f}{2}} \,
  \frac{\exp\Bigl ( -\frac{(\xi-1)^2}{2 f}- \frac{f}{2}\ln^2 \xi\Bigr)}{\Gamma(\xi)}
  .
\end{gather}
Finally, using Eq.~(\ref{DOS2}), we find
\begin{equation}
\label{DOS8}
  D(\eps) = \frac{n_L}{E_0} S_f(\xi),
\end{equation}
where $S_f(\xi)$ is defined in Eq.~\eqref{Sf}.
Equation~\eqref{DOS8} gives the result \eqref{DOSRES} in the region
$ 0 \leq E \ll E_0 e^{-1/f}$.

\begin{figure}
\centerline{\includegraphics[width=240pt,height=210pt]{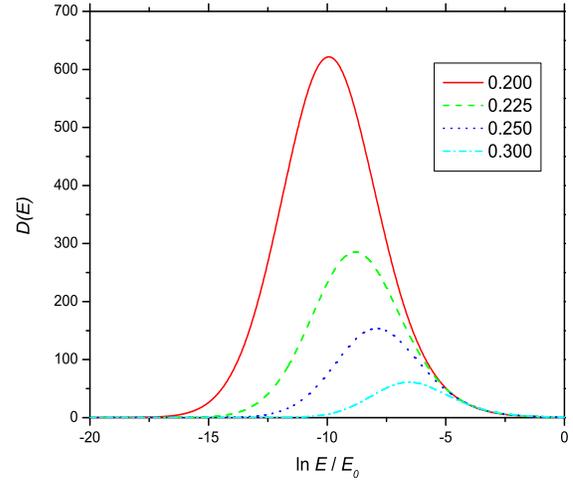}}
\caption{Fig.~\protect\ref{FIG2}. The density of states $D(E)$ in
units of $n_L/\pi E_0$ as a function of $\ln E/E_0$ for different
values of impurity concentration $f$.} \label{FIG2}
\end{figure}

The whole profile of $D(E)$ for $\eps \ll 1$  can
be obtained by numerical evaluation of Eqs.~\eqref{DOS2} and
\eqref{F12}. The density of states numerically calculated for
several values of the impurity concentration $f$ is presented in
Fig.~\ref{FIG2}.

{\bf 5. Conclusion.}

In conclusion, we evaluated the density of states of a
two-dimensional electron gas in the presence of the strong
magnetic field and impurities. The fact that impurities are
situated at different distances from two-dimensional electron gas
leads to the dramatic change of the density of states in the case
of small impurity concentration compared to the case when all
impurities are situated at the same distance from the 2DEG.

Using the exact result of Ref.~\cite{BGI} we obtained the
density of states in the whole energy range for the case
of the wave function with the asymptotic behavior (\ref{varphi}).
The density of states
vanishes at the position of the unperturbed Landau level
and has a maximum at an exponentially small energy (\ref{Emax}).
The major part of the states are localized by single impurities
in accordance with findings of Ref.~\cite{DGO}.

The functional form of the result will be different for asymptotic
behavior of $\varphi(z)$ differing from the simple exponential
decay (\ref{varphi}). However, the qualitative structure of
the density of states is supposed to be preserved.


We acknowledge useful discussions with M.V. Feigelman, S.V.
Iordansky, A.S. Iosselevich, D. Lyubshin, and P.M. Ostrovsky. We
are grateful to A.M. Dyugaev, P.D. Grigoriev and Yu.N. Ovchinnikov
for bringing their paper to us prior to publication. We thank
Forschungszentrum J\"ulich (Landau Scholarship) (I.~S.~B.),
Dynasty Foundation, ICFPM, RFBR under grant 01-02-17759, and
Russian Ministry of Science (M.~A.~S.) for financial support.


\end{document}